\documentclass[12pt,preprint]{aastex} 
\usepackage{emulateapj5}
\usepackage{epsfig} 
\newcommand{\chandra}{{\it{Chandra}}}
\newcommand{\rosat}{{\it ROSAT}} 
\newcommand{\asca}{{\it ASCA}}
\newcommand{\xmm}{{\it XMM}} 

\newcommand{\beppo}{{\it BeppoSAX}}
\newcommand{\mdot}{\dot{M}} 
\newcommand{\msunyr}{{\rm M}_{\sun}~{\rm yr}^{-1}} 
\newcommand{\fx}{ergs s$^{-1}$ cm$^{-2}$}
\newcommand{\lxh}{$h_{70}^{-2}$ ergs s$^{-1}$}
\newcommand{\hkpc}{$h_{70}^{-1}$~kpc}
\newcommand{\hmpc}{$h_{70}^{-1}$~Mpc} 
\newcommand{\zsun}{$Z_{\sun}$}
\newcommand{\myfigure}{figurehere} 

\submitted{Received 2002 January 24; Accepted 2002 May 22}
\shorttitle{\chandra{} observations of A2029} 
\shortauthors{Lewis et al.}

\begin{document}

\title{\chandra{} Observations of Abell 2029: No Cooling Flow and a
Steep Abundance Gradient}

\author{Aaron D. Lewis\altaffilmark{1,2}, John T.
Stocke\altaffilmark{2}, David A. Buote\altaffilmark{1}}
\email{lewisa@uci.edu, stocke@casa.colorado.edu, buote@uci.edu}

\altaffiltext{1}{University of California, Irvine, Department of Physics
and Astronomy, 4171 Frederick Reines Hall, Irvine, CA, 92697-4575}
\altaffiltext{2}{Center for Astrophysics and Space Astronomy, University
of Colorado, 389 UCB,  Boulder, CO 80309}

\received{}

\accepted{}

\journalid{}{}

\articleid{}{}

\begin{abstract} We have obtained high spatial resolution temperature
and abundance profiles for the galaxy cluster Abell 2029 with the
\chandra{} ACIS-S instrument. Our observations reveal that the spectra
are well-fit by a single-phase gas in each annulus. While the
temperature of the intracluster medium drops from $\sim9$~keV at
3\arcmin{} (260~\hkpc) to $3$~keV in the central $5\arcsec$ (5~\hkpc) of
the cluster, there is no evidence for gas emitting at temperatures below
3~keV. The addition of a cooling flow component does not improve the
fits, despite previous claims for a massive cooling flow. There is also
no evidence for excess absorption above the Galactic N$_H$ value. We
also observe a steep Fe abundance gradient, such that $Z_{Fe} \gtrsim
2$~\zsun (assuming meteoritic solar Fe abundance) in the core,
consistent with significant enrichment from Fe-rich Type Ia supernovae
in the cD galaxy. The Fe abundance drops to $\approx 0.5$~\zsun{} at
3\arcmin{} (260~\hkpc), in good agreement with previous \beppo{}
measurements. The high resolution image reveals neither a strong central
point source, nor any filamentary structure related to a cooling flow or
a merger. The absence of a strong merger signature argues against the
creation of the wide-angle-tail radio source morphology in a merger
event. \end{abstract}


\section{Introduction\label{sec_intro}} 

Abell 2029 (A2029 hereafter) is a richness class II, Bautz-Morgan type I
nearby cluster of galaxies at a redshift of $z=0.0767$ \citep{abe89}. It
has a very large cD galaxy (UGC 9752 = IC 1101) with a stellar envelope
extending well beyond 600~\hkpc{} \citep{uso91}, and is one of the most
optically regular rich clusters known \citep{dre78}. It has been well
studied optically \citep[see e.g.,][]{dre81,joh87}, and can be
characterized as a compact, relaxed, cD-galaxy dominated system, with no
emission-line emitting galaxies within 600~\hkpc{} of its core
\citep{dal00}. Throughout this Letter, we assume a cosmology of
H$_0=70$~$h_{70}$~km~s$^{-1}$~Mpc$^{-1}$, $\Omega_{matter}=0.3$, and
$\Lambda=0.7$, implying a luminosity distance to A2029 of 347~\hmpc{}
and a scale of 1.45 kpc~arcsec$^{-1}$.

A2029 has been previously studied in the X-rays using various
instruments:  the \rosat{} HRI \citep{sar92}, \rosat{} PSPC and \asca{}
\citep{buo96,sar98,whi00}, and \beppo{} \citep{mol99,irw00}. It is
one of the brightest X-ray clusters in the sky ($F_X = 7.5 \times
10^{-11}$\fx{} in the 2-10 keV band, \citealt{dav93}), corresponding to
$L_X (2-10$keV)$ =1.1 \times 10^{45}$~\lxh{}. From \asca{} spectra,
\citet{mol99} find an intracluster medium (ICM) temperature in the
central 2\arcmin{} of $T_X = 8.3\pm0.2$~keV declining to $\sim5$~keV at
a radius of 10\arcmin{} (972~\hkpc).

Previous imaging and spectral studies have inferred a large cooling flow
with $\mdot\simeq 400-600$~$\msunyr$ \citep{edg92,fabi94,arn88,per98},
\citep[but c.f.,][]{whi00}. However, evidence that it is not a cooling
flow includes no detection of H$\alpha$ or \ion{O}{2} emission in the cD
galaxy, and no blue stellar colors in the core \citep{mcn89}. The cD
galaxy also hosts a wide-angle-tail (WAT) radio source (PKS 1509+49 =
4C+06.53), which is highly unusual in a cooling flow system
\citep{bur90}. Simulations show that WATs may be the signatures of
ongoing cluster mergers; these simulated systems also exhibit hot shock
fronts \citep{lok95,roe96}.

Obviously the relaxed cooling flow and merger/hot shock scenarios are
mutually exclusive, and for this reason we have obtained high spatial
resolution imaging spectroscopy with the \chandra{} observatory in order
to 
\begin{\myfigure}
\centerline{\epsfig{file=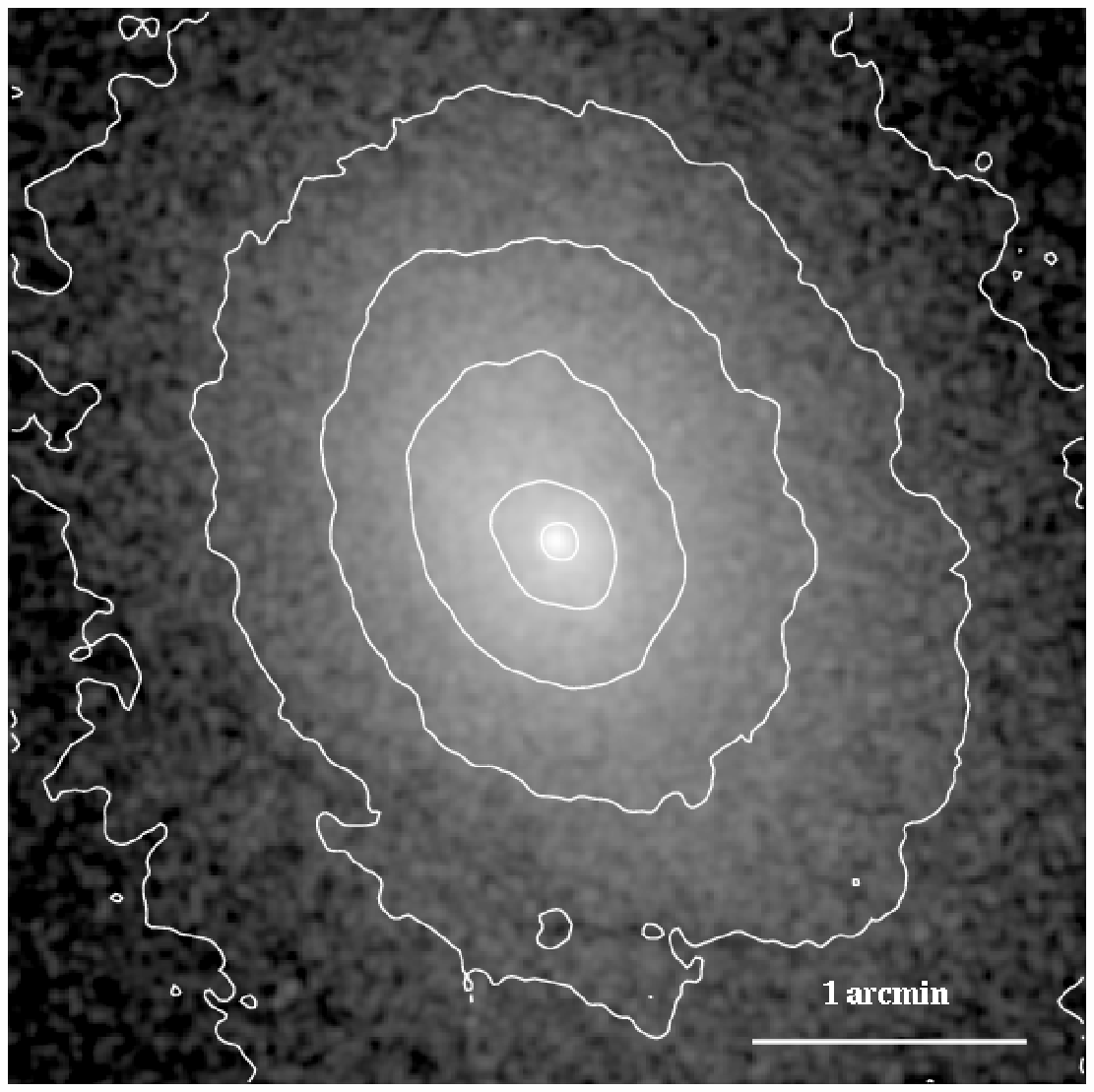, scale=0.50}}
\caption{\chandra{} ACIS-S image of Abell 2029. The image is 4$\arcmin$
(348 \hkpc) on a side and has been smoothed with a Gaussian of $\sigma =
2\arcsec$. North is up and East is left. Logarithmically spaced contours
are overlaid to indicate morphology.}
\label{fig_image}
\end{\myfigure}
\smallskip
\noindent 
investigate this enigmatic object. In this Letter we describe the
global temperature structure, resolved metal abundances, as well as
highly spatially-resolved spectral properties of the central regions. A
future publication will measure the dark matter profile.

We note that the authors of most X-ray spectral analyses to date have
used ``photospheric'' Fe abundances. Following the suggestion of
\citet{ish97}, we use the abundance measurements of \citet{and89} for
spectral fitting, except for Fe where we use the correct ``meteoritic''
value, Fe/H $ = 3.24 \times 10^{-5}$ by number. This results in Fe
abundance values a factor 1.44 times larger than the ``photospheric''
values. We will note this factor when comparing to other work.

\section{Observations and Data Reduction\label{sec_obs}}

A2029 was observed with the \chandra{} observatory on 2000 12 April for
a deadtime-corrected exposure of 19.8 ksec.  The cluster was centered on
the backside-illuminated S3 chip, operating at a temperature of -120 C.
The data were re-processed using CIAO 2.2.1 and version 2.7 of the
\chandra{} calibration database (CALDB). We mitigated the effects of
charge transfer inefficiency using the ACISCtiCorrector.1.37
software\footnote{Available from the \chandra{} contributed software
page at \url{http://asc.harvard.edu/cont-soft/soft-exchange.html}}
\citep{tow00}. The data were screened for energy (0.3 to 8.0 keV), grade
(ASCA 02346), and status (0). We examined the lightcurve for possible
flaring of the background, and removed 365 seconds of suspect exposure.
We then scaled and subtracted the available source-free extragalactic
sky background maps\footnote{The background maps have been CTI-corrected
in exactly the same manner as the source data.} using the make\_acisbg
software created by Maxim Markevitch \citep[see, e.g.][]{mar00}.

In Figure \ref{fig_image} we present the smoothed, background-subtracted ACIS-S
image of A2029 in the energy range $0.3-8.0$ keV. The image reveals A2029 to be
quite regular; although there are some position-angle dependent variations
within a $1\arcmin$ radius, we find no evidence of bright shock fronts,
filaments, or bright point sources. The X-ray emission is somewhat elliptical,
with a position angle of $\sim25\arcdeg$ (North to East), consistent with the
\rosat{} PSPC image analysis of \citet{buo96}.

\section{Spatially Resolved Spectral Properties \label{sec_space}}

To analyze the spectrum of the hot ICM, we extracted spectra from eleven
concentric annuli centered on the peak of the X-ray emission (coincident
with the BCG at RA = $15^h 10^m 56.^s2$, Decl. = $+ 05\arcdeg 44\arcmin
44\arcsec$, J2000). The annuli were of varying widths providing
approximately 10,000 counts per annulus in the energy range $0.3-8.0$
keV, except that to investigate the core with maximum spatial
resolution, we chose a central region $5\arcsec$ in radius, containing
approximately 5,000 counts.

\subsection{Radial Temperature Profile\label{subsec_radt}}

Using {\sc xspec}, we fit the extracted spectra with the {\sc apec}
plasma emission model\footnote{We have also performed fits with the {\sc
mekal} model. We find no statistical differences in temperature or
abundance measurements.} absorbed by neutral hydrogen. We adopt the
weighted average Galactic value of N$_H = 3.14 \times 10^{20}$~cm$^{-2}$
obtained from the W3N$_H$ HEASARC tool.   
The Fe abundance was allowed to
be a free parameter, with all other elements tied to Fe in their solar
ratios.

\begin{\myfigure}
\centerline{\epsfig{file=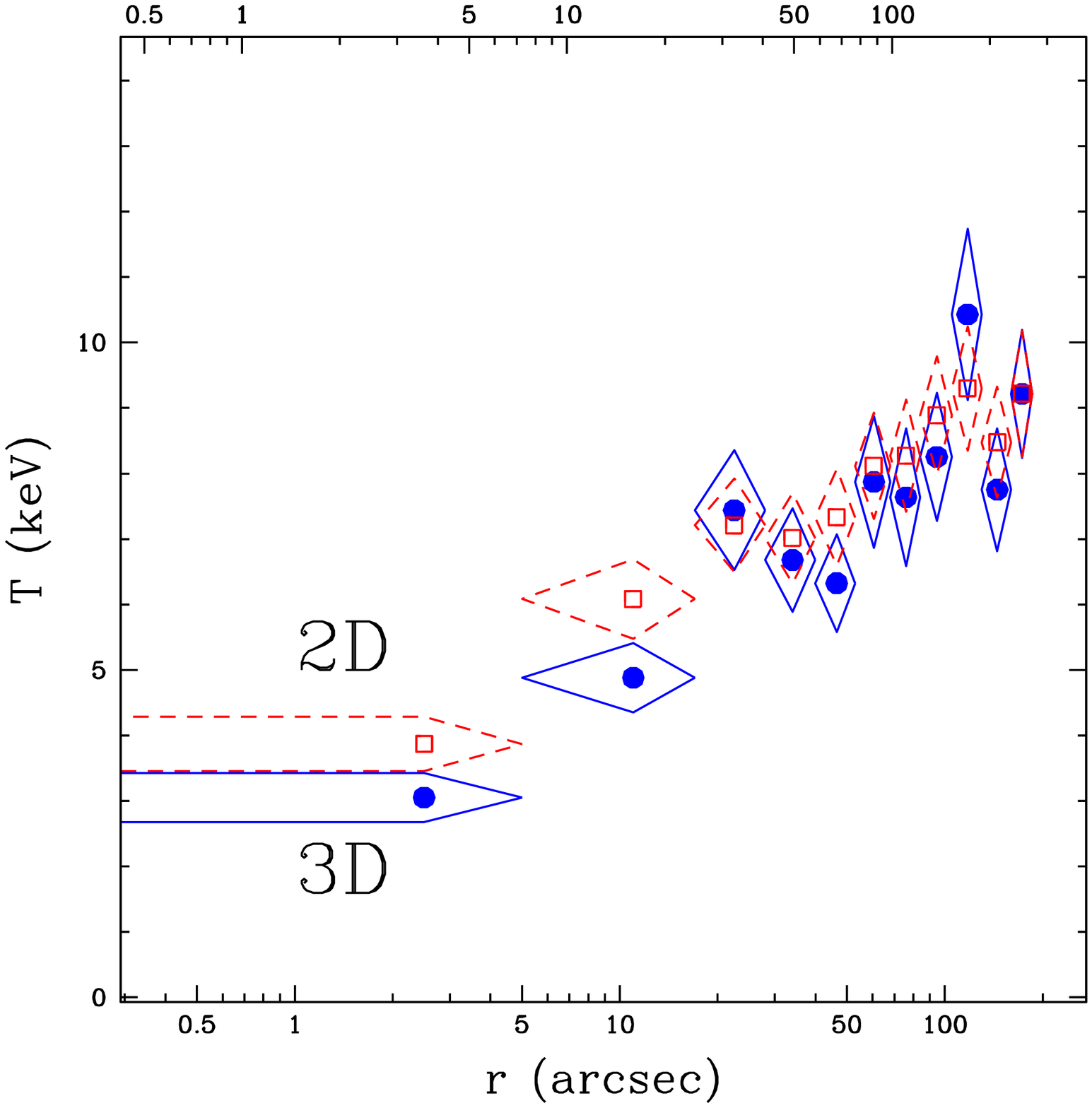, scale=0.33}}
\caption{\chandra{} radial temperature profile of A2029. Solid lines (and filled
circles) are the results of the 3D spectral deprojection. Dotted lines (and
open squares) are the 2D analysis. Horizontal axis shows both arcsec ({\it lower
axis}) and \hkpc ({\it upper axis}) units.}
\label{fig_t}
\end{\myfigure}
\smallskip
 
To properly recover the three-dimensional properties of the
X-ray emitting ICM, we have performed a spectral deprojection analysis
using the {\sc xdeproj} code of \citet{buo00c}. We start at the outside
working our way to the core in an ``onion-peeling'' method which
accounts for the cumulative projected emission from the outer annuli,
obtaining temperatures, densities, abundances, and any other desired
parameters. For details of our deprojection technique see \citet{buo00c}
and \citet{P7}. 
To estimate the uncertainties on the fitted parameters we simulated
spectra for each annulus using the best-fitting models and fit the
simulated spectra in exactly the same manner as the actual data. From
100 Monte Carlo simulations we compute the standard deviation for each
free parameter which we quote as the ``1$\sigma$'' error. 
The reference model we have used is {\sc apec} with deprojection,
denoted {\sc apec} (3D).
For comparison, we have also performed a standard two-dimensional
analysis with no deprojection, denoted {\sc apec} (2D).

%
In Figure \ref{fig_t} we show the
radial temperature profile of A2029 derived from the 3D analysis ({\it
solid lines} and filled circles) and the 2D analysis ({\it dotted lines}
and open squares). Figure \ref{fig_t} shows an obvious drop in
temperature in the core, with increasing values at larger radii (our
final bin is between 160-186 arcsec, $2.7-3.1\arcmin$, or
232-270~\hkpc). The 3D model shows a more dramatic drop in the center,
the signal of which is masked in the 2D fits as hotter gas from larger
radii is observed in projection over the cooler core. We observe the gas
temperature to be approximately 8-10 keV at 100-200 arcseconds. The gas
drops to a minimum temperature of $3.1\pm0.4$~keV in the inner
$5\arcsec$. The data suggest a continuing increase in temperature,
though the profile is possibly flattening at the limit of our
measurements. We note that the sharp bin-to-bin fluctuations seen for
the 3D model in the four outermost annuli are due to the nature of the
3D deprojection algorithm, and should not be considered a physical
feature. This effect can be avoided by regularizing (in effect
smoothing) the parameters, which we do not perform for this Letter (but
see \citealt{buo00c} for details of the procedure). The 3D fluctuations
in the outer bins vary about the 2D temperature values in these bins
without significant bias. This is expected because at large radii 
the gas is nearly isothermal and also suffers the least from projection
effects, such that the 2D profile can be used to verify the 3D results in these
annuli. The quality of the fits is only marginally improved using the 3D vs. 2D
method, with the most improvement occurring in the inner 2 annuli where we also
observe the largest change in temperature.

The formal quality of the spectral fits is quite good ($\chi^2$/dof =
134.02/111 in the central annulus, for the reference model), but there exist
significant (10-15\%) residuals below 0.5 keV, where the data are in excess of
the model. At these energies we have either under-subtracted the background or
there are calibration issues outstanding. In the outer annuli where the
observations are background dominated (especially at high energy), the
background estimate seems secure and we are inclined to doubt the calibration
at E $< 0.5$~keV. Between 0.65 and 0.75 keV, the data fall below the model in
the inner three annuli, but this `absorption feature' is much sharper than the
spectral resolution of the S3 chip and its nature is not clear. These features
can be see in Fig. \ref{fig_spec}, wherein we present the spectrum and
residuals from the fit in Annulus 2 (between 5 and 17\arcsec).

Allowing the neutral hydrogen column density to be a free parameter in
the fits results in N$_H$ values less than half the Galactic value in
the outer regions which climb to slightly higher than Galactic in the
central bin.
As a test, we have performed an additional set of fits eliminating all
energies below 0.7 keV ({\sc apec} (3D-0.7) $\chi^2$/dof = 87.10/91). We
have repeated this exercise with all the other models presented in this
Letter. We find that the removal of low energy residuals does not change
either the temperature or abundance measurements beyond their errors.
The rise in N$_H$ in the center would appear to accommodate the
`absorption feature' near 0.6 keV in the central bin, but elimination of
energies below 0.7 keV merely weakens the constraint on N$_H$ without
lowering the fitted value. In either case, there is almost no
improvement in the fit quality in the inner 2 annuli. Freeing N$_H$ does
result in a significant improvement in the fit quality for annuli 3-11.
As expected, eliminating energies below 0.7 keV in the fits obtains
higher (but still sub-Galactic) N$_H$ values. Although it is possible
there is a decrement in actual Galactic N$_H$ along the line of sight to
A2029, this would not be consistent with any previous observations (at a
variety of wavelengths).
%

\subsection{Fe abundance profile}

In Figure \ref{fig_z} we show the radial metal abundance profile of
A2029 from the {\sc apec} (3D) model fits. Ratio of Fe to its solar
(meteoritic) value is shown. We observe subsolar (approximately 0.5
\zsun) abundances outside of 20\arcsec, and a very sharp increase in the
core to $\sim2$ \zsun. The abundances of other elements (S, Si, Mg and
O) are not constrained to better than $\sim50\%$ except in a few central
bins where the emission lines are more pronounced relative to the lower

\begin{\myfigure}
\centerline{\epsfig{file=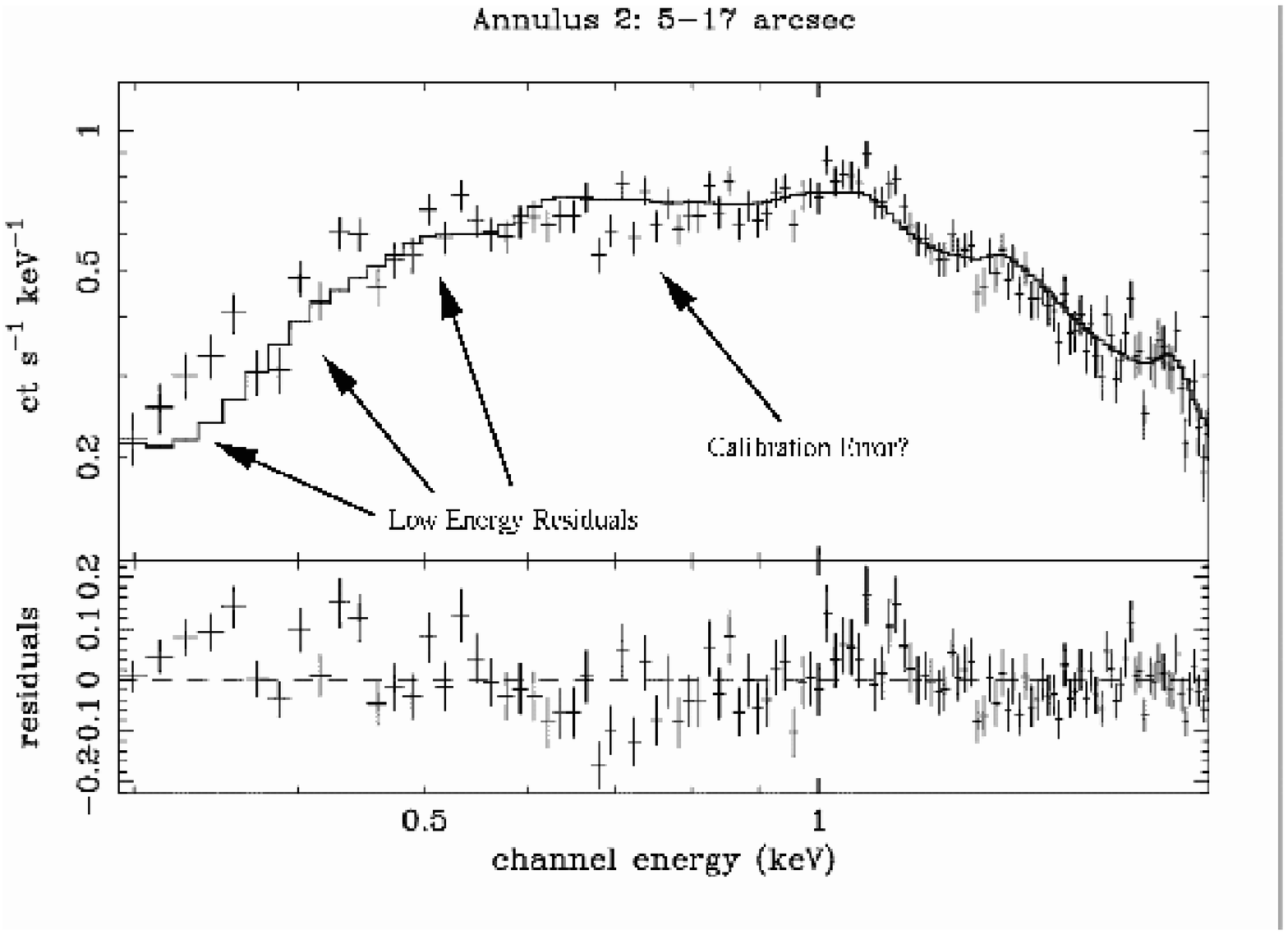, scale=0.30}}
\caption{Spectrum in Annulus 2 (5-17\arcsec) between 0.3 and 2.0 keV.
Fractional residuals shown at bottom. Model is {\sc apec} (3D), with Galactic
N$_H$ absorption. Note that we do not see these residual features at larger radii.}
\label{fig_spec}
\end{\myfigure}
\smallskip
\begin{\myfigure}
\centerline{\epsfig{file=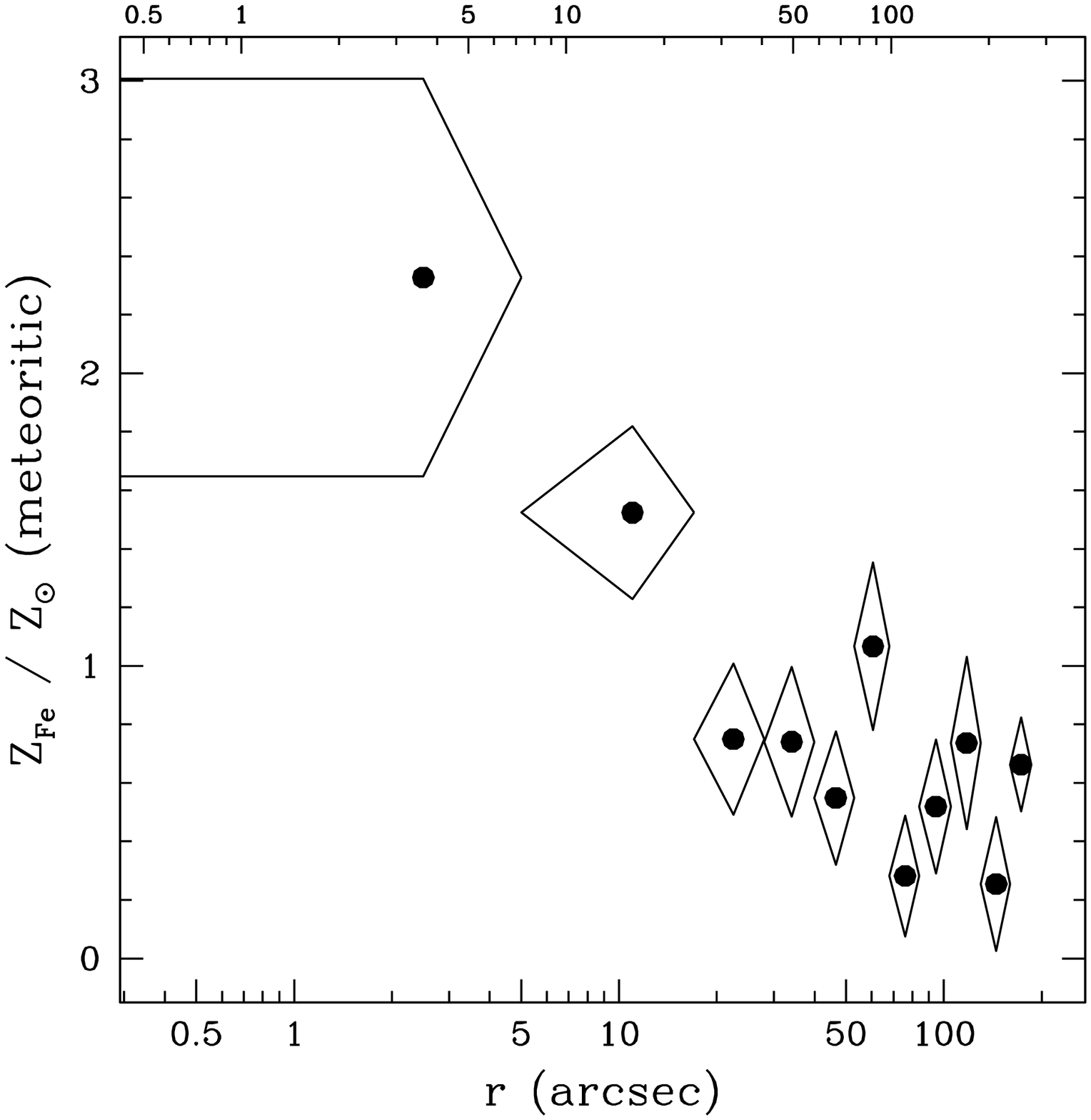, scale=0.30}}
\caption{\chandra{} deprojected radial Fe abundance profile of A2029. 
Horizontal axis shows both arcsec ({\it lower axis}) and kpc ({\it upper
axis}) units.}
\label{fig_z}
\end{\myfigure}
\smallskip
\noindent
temperature ICM. As we are unable to place significant constraints on
any interesting abundance ratios we do not present these data. We note
that allowing additional element abundances to vary does not effect the
Fe abundances beyond the errors shown in Figure \ref{fig_z}.

Our measurements are entirely consistent with the $\sim$arcmin spatial
resolution \beppo{} results of \citet{mol99} and \citet{irw01},
who measured $Z_{Fe} \sim0.5$~\zsun{} (0.72~\zsun{} in meteoritic units)
within the central 2\arcmin, declining to $\sim0.2$ (0.29) at larger
radii ($\sim10\arcmin$). An emission-weighted average of our
best-fitting \chandra{} Fe abundance measurements in bins 1-9
($0-2.2\arcmin$) yields $Z_{Fe} = 0.77$~\zsun{} (meteoritic). We also
see a decline to values consistent with 0.35~\zsun{} (within our errors)
for our outermost data points which extend to 3\arcmin.
%
%
%
%

\section{Analysis of the Core\label{sec_core}}

The unsurpassed spatial resolution of \chandra{} allows us to
investigate the core of A2029 with high S/N in the central 5\arcsec{}
region. The observation of A2029 was pointed 1.3\arcmin{} from the
ACIS-S aimpoint (the cluster center was placed 1.2\arcmin{} south, and
30\arcsec{} east of the aimpoint to place more extended emission on the
S3 chip), which should result in an effective PSF of $\lesssim
1.5\arcsec$ FWHM (it is within the region of the detector where the
encircled energy within 1\arcsec{} is $\geq 50\%$\footnote{see the
\chandra{} Proposers' Observatory Guide Rev 3.0.}).

\subsection{Alternative Spectral Models\label{subsec_alt}}

From our single-temperature fit we obtain $T_X = 3.1\pm0.4$~keV, and a
good fit to several strong emission features across the entire spectrum,
except for the low energy absorption feature near 0.6 keV. Although the
quality of the fit in this annulus is acceptable (134.02/111
$\chi^2$/dof), we have investigated additional or alternative spectral
models.

We have added to the first three annuli in turn: a second temperature
component ({\sc apec+apec} (3D) $\chi^2$/dof = 132.60/108), a constant
pressure cooling flow, with the material cooling from the temperature of
the {\sc apec} model ({\sc apec}+CF (3D) $\chi^2$/dof = 133.29/109), and
a power law model ({\sc apec}+POW (3D) $\chi^2$/dof = 133.62/109). Each
model was absorbed by N$_H$, which was fixed to the Galactic value for
all models except CF. None of these models provided a significant
improvement to the fits. The cooling flow fit yields a cooling rate of
$0.0\pm0.7\msunyr$, similar to other recent XMM-Newton analyses \citep[see,
e.g.,][]{kaa01,boh02}.

\subsection{Constraints on a Central Point Source}

In Figure \ref{fig_sb} we show the azimuthally averaged surface
brightness profile of A2029 obtained from counts in the $0.3-8.0$~kev
band. We extracted counts from the background-subtracted image in annuli
1.5\arcsec{} wide out to a radius of 186\arcsec{}. We then subtracted an
in-field background taken from a region outside the extraction area
still on the S3 chip (the southwest corner). 
For the present analysis of
the core, the impact of an exposure map is negligible, and we have not 
performed such a correction.
\begin{\myfigure}
\centerline{\epsfig{file=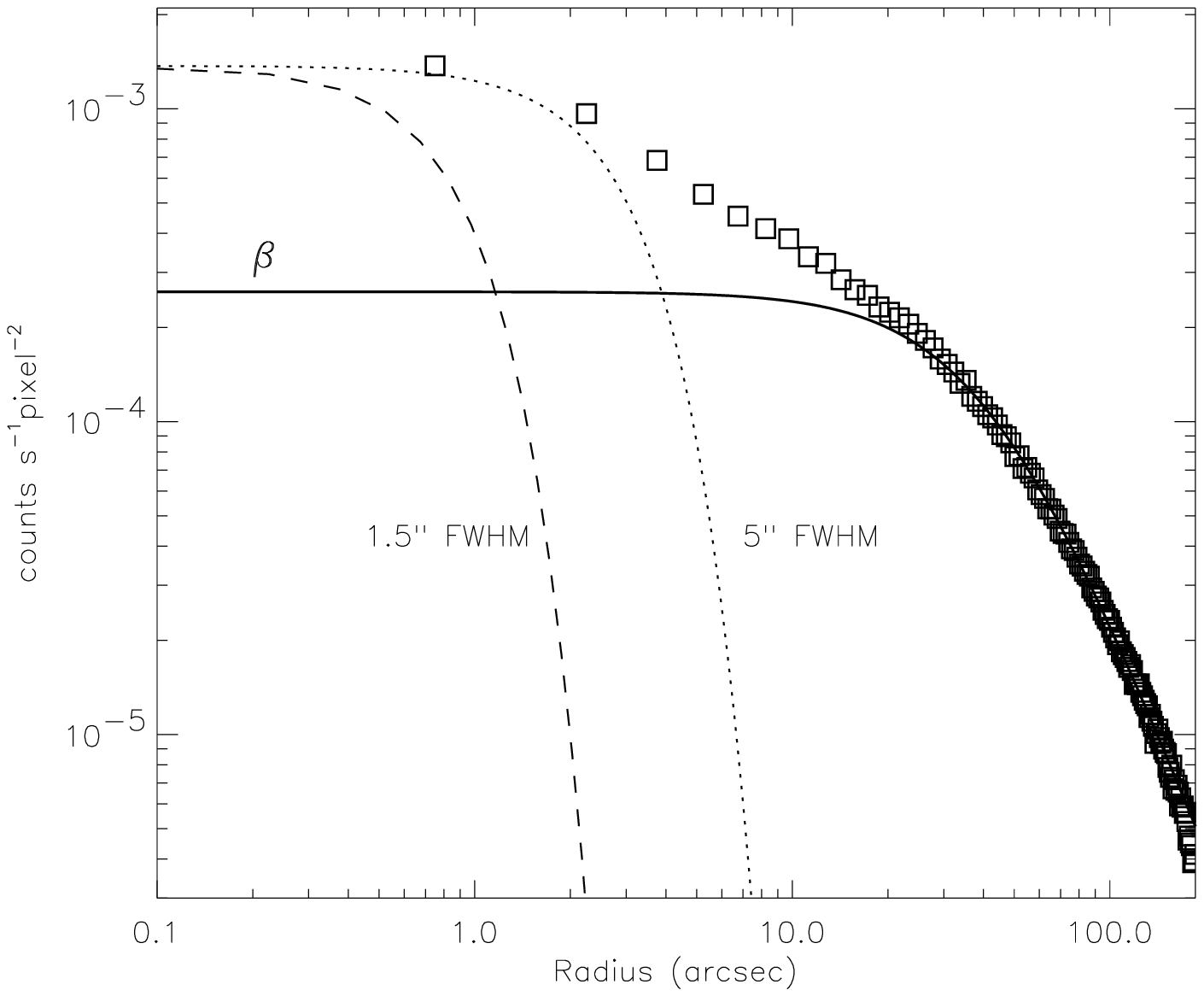, scale=0.5}}
\caption{\chandra{} surface brightness profile of A2029 in the
$0.3-8.0$~keV band ({\it open squares}). Overlaid are two Gaussians of
FWHM 1.5 ({\it dashed curve}) and 5\arcsec ({\it dotted curve}),
respectively, as well as a $\beta$-model fit ({\it solid curve}), with
$\beta=0.60$, $r_{core}=42\arcsec$.}
\label{fig_sb}\end{\myfigure}
\smallskip

The profile is smooth, and well constrained to the limit of our
observations. There is a significant peak in the profile at $\lesssim
20\arcsec$. This peak is wider than a Gaussian of 5\arcsec{} FWHM
(overlaid as a dotted line), and thus unambiguously resolved (for
reference we have also overlaid a Gaussian of 1.5\arcsec{} FWHM [{\it
dashed line}] representing an upper limit to the ACIS PSF). We also show
a standard $\beta$-model 
fit to the data omitting the inner 20\arcsec{}
to obtain an acceptable fit. Because we see no spectral evidence for
either a power-law source, or a cooling flow, we suggest that the peak
may represent a cusp in the gravitational potential due to the mass of
the cD galaxy. 

\section{Discussion} We have analyzed high spatial resolution \chandra{}
spectra of the A2029 cluster of galaxies. We observe both strong temperature
and abundance gradients within the central arcmin, with $T_X$ dropping to
$\sim3$~keV, and $Z_{Fe}$ rising to $\approx2$~\zsun. We find no evidence for
excess absorption in the core, though there are some remaining residuals in the
low energy ($\le0.7$~keV) region of the spectrum which are not yet understood.
Additional model components (including a cooling flow) are not required, and do
not improve the fits. Thus, A2029 bears no evidence of multi-phase gas,
refuting earlier claims of $\mdot \sim 400-600 \msunyr$ and thus is similar to,
although more extreme than, many other recent \chandra{} and \xmm{}
observations \citep[see][for a review]{mcn02}. From the image, we find no
evidence of a strong shock or major cluster merger, and thus a lack of support
for the \citet{lok95} model for producing WAT radio sources in cluster merger
events. There is a small peak in the inner 10\arcsec{} of the surface brightness
profile, indicating a second component in excess of a standard $\beta$-model.

The super-solar Fe abundance measurements in the core of A2029 are
suggestive of enrichment by metals created by Fe-rich SN Ia and ejected
from the cD and other galaxies in the core. Since the gaseous system is
relaxed and fairly undisturbed (no recent mergers), this gradient could
have been maintained to the present epoch from an earlier time. This
also agrees with the analysis of \citet{deg01} (although we now resolve
the enrichment at $<2\arcmin$), who found that the abundance gradient
matched the stellar light profile of the galaxies in the cluster core.

\acknowledgments

This work was supported by \chandra{} grant G00-1021X. ADL thanks Beth
Lewis for her ongoing support.


\begin{thebibliography}{32}
\expandafter\ifx\csname natexlab\endcsname\relax\def\natexlab#1{#1}\fi

\bibitem[{{Abell} {et~al.}(1989){Abell}, {Corwin}, \& {Olowin}}]{abe89}
{Abell}, G.~O., {Corwin}, H.~G., \& {Olowin}, R.~P. 1989, \apjs, 70, 1

\bibitem[{{Anders} \& {Grevesse}(1989)}]{and89}
{Anders}, E. \& {Grevesse}, N. 1989, \gca, 53, 197

\bibitem[{{Arnaud}(1988)}]{arn88}
{Arnaud}, K.~A. 1988, in NATO ASIC Proc. 229: Cooling Flows in Clusters and
  Galaxies, 31--40

\bibitem[{{B{\" o}hringer} {et~al.}(2002){B{\" o}hringer}, {Matsushita},
  {Churazov}, {Ikebe}, \& {Chen}}]{boh02}
{B{\" o}hringer}, H., {Matsushita}, K., {Churazov}, E., {Ikebe}, Y., \& {Chen},
  Y. 2002, \aap, 382, 804

\bibitem[{{Buote}(2000)}]{buo00c}
{Buote}, D.~A. 2000, \apj, 539, 172

\bibitem[{{Buote} \& {Canizares}(1996)}]{buo96}
{Buote}, D.~A. \& {Canizares}, C.~R. 1996, \apj, 457, 565

\bibitem[{{Buote} {et~al.}(2002){Buote}, {Lewis}, {Brighenti}, \&
  {Mathews}}]{P7}
{Buote}, D.~A., {Lewis}, A.~D., {Brighenti}, F., \& {Mathews}, W.~G. 2002,
  ~ApJ, submitted

\bibitem[{{Burns}(1990)}]{bur90}
{Burns}, J.~O. 1990, \aj, 99, 14

\bibitem[{{Dale} \& {Uson}(2000)}]{dal00}
{Dale}, D.~A. \& {Uson}, J.~M. 2000, \aj, 120, 552

\bibitem[{{David} {et~al.}(1993){David}, {Slyz}, {Jones}, {Forman}, {Vrtilek},
  \& {Arnaud}}]{dav93}
{David}, L.~P., {Slyz}, A., {Jones}, C., {Forman}, W., {Vrtilek}, S.~D., \&
  {Arnaud}, K.~A. 1993, \apj, 412, 479

\bibitem[{{De Grandi} \& {Molendi}(2001)}]{deg01}
{De Grandi}, S. \& {Molendi}, S. 2001, \apj, 551, 153

\bibitem[{{Dressler}(1978)}]{dre78}
{Dressler}, A. 1978, \apj, 226, 55

\bibitem[{{Dressler}(1981)}]{dre81}
---. 1981, \apj, 243, 26

\bibitem[{{Edge} {et~al.}(1992){Edge}, {Stewart}, \& {Fabian}}]{edg92}
{Edge}, A.~C., {Stewart}, G.~C., \& {Fabian}, A.~C. 1992, \mnras, 258, 177

\bibitem[{{Fabian}(1994)}]{fabi94}
{Fabian}, A.~C. 1994, \araa, 32, 277

\bibitem[{{Irwin} \& {Bregman}(2000)}]{irw00}
{Irwin}, J.~A. \& {Bregman}, J.~N. 2000, \apj, 538, 543

\bibitem[{{Irwin} \& {Bregman}(2001)}]{irw01}
---. 2001, \apj, 546, 150

\bibitem[{{Ishimaru} \& {Arimoto}(1997)}]{ish97}
{Ishimaru}, Y. \& {Arimoto}, N. 1997, \pasj, 49, 1

\bibitem[{{Johnstone} {et~al.}(1987){Johnstone}, {Fabian}, \& {Nulsen}}]{joh87}
{Johnstone}, R.~M., {Fabian}, A.~C., \& {Nulsen}, P.~E.~J. 1987, \mnras, 224,
  75

\bibitem[{{Kaastra} {et~al.}(2001){Kaastra}, {Ferrigno}, {Tamura}, {Paerels},
  {Peterson}, \& {Mittaz}}]{kaa01}
{Kaastra}, J.~S., {Ferrigno}, C., {Tamura}, T., {Paerels}, F.~B.~S.,
  {Peterson}, J.~R., \& {Mittaz}, J.~P.~D. 2001, \aap, 365, L99

\bibitem[{{Loken} {et~al.}(1995){Loken}, {Roettiger}, {Burns}, \&
  {Norman}}]{lok95}
{Loken}, C., {Roettiger}, K., {Burns}, J.~O., \& {Norman}, M. 1995, \apj, 445,
  80

\bibitem[{{Markevitch} {et~al.}(2000){Markevitch}, {Ponman}, {Nulsen}, {Bautz},
  {Burke}, {David}, {Davis}, {Donnelly}, {Forman}, {Jones}, {Kaastra},
  {Kellogg}, {Kim}, {Kolodziejczak}, {Mazzotta}, {Pagliaro}, {Patel}, {Van
  Speybroeck}, {Vikhlinin}, {Vrtilek}, {Wise}, \& {Zhao}}]{mar00}
{Markevitch}, M., {Ponman}, T.~J., {Nulsen}, P.~E.~J., {Bautz}, M.~W., {Burke},
  D.~J., {David}, L.~P., {Davis}, D., {Donnelly}, R.~H., {Forman}, W.~R.,
  {Jones}, C., {Kaastra}, J., {Kellogg}, E., {Kim}, D.-W., {Kolodziejczak}, J.,
  {Mazzotta}, P., {Pagliaro}, A., {Patel}, S., {Van Speybroeck}, L.,
  {Vikhlinin}, A., {Vrtilek}, J., {Wise}, M., \& {Zhao}, P. 2000, \apj, 541,
  542

\bibitem[{{McNamara}(2002)}]{mcn02}
{McNamara}, B.~R. 2002, in ASP Conf. Ser., X-rays at Sharp Focus: Chandra
  Science Symposium, ed. S. Vrtilek, E. M. Schlegel, \& L. Kuhi (San Francisco:
  ASP), astro-ph/0202199

\bibitem[{{McNamara} \& {O'Connell}(1989)}]{mcn89}
{McNamara}, B.~R. \& {O'Connell}, R.~W. 1989, \aj, 98, 2018

\bibitem[{{Molendi} \& {De Grandi}(1999)}]{mol99}
{Molendi}, S. \& {De Grandi}, S. 1999, \aap, 351, L41

\bibitem[{{Peres} {et~al.}(1998){Peres}, {Fabian}, {Edge}, {Allen},
  {Johnstone}, \& {White}}]{per98}
{Peres}, C.~B., {Fabian}, A.~C., {Edge}, A.~C., {Allen}, S.~W., {Johnstone},
  R.~M., \& {White}, D.~A. 1998, \mnras, 298, 416

\bibitem[{{Roettiger} {et~al.}(1996){Roettiger}, {Burns}, \& {Loken}}]{roe96}
{Roettiger}, K., {Burns}, J.~O., \& {Loken}, C. 1996, \apj, 473, 651

\bibitem[{{Sarazin} {et~al.}(1992){Sarazin}, {O'Connell}, \&
  {McNamara}}]{sar92}
{Sarazin}, C.~L., {O'Connell}, R.~W., \& {McNamara}, B.~R. 1992, \apjl, 389,
  L59

\bibitem[{{Sarazin} {et~al.}(1998){Sarazin}, {Wise}, \& {Markevitch}}]{sar98}
{Sarazin}, C.~L., {Wise}, M.~W., \& {Markevitch}, M.~L. 1998, \apj, 498, 606

\bibitem[{{Townsley} {et~al.}(2000){Townsley}, {Broos}, {Garmire}, \&
  {Nousek}}]{tow00}
{Townsley}, L.~K., {Broos}, P.~S., {Garmire}, G.~P., \& {Nousek}, J.~A. 2000,
  \apjl, 534, L139

\bibitem[{{Uson} {et~al.}(1991){Uson}, {Boughn}, \& {Kuhn}}]{uso91}
{Uson}, J.~M., {Boughn}, S.~P., \& {Kuhn}, J.~R. 1991, \apj, 369, 46

\bibitem[{{White}(2000)}]{whi00}
{White}, D.~A. 2000, \mnras, 312, 663

\end{thebibliography}

\end{document}